\documentclass[english,aps,prx,reprint,superscriptaddress]{revtex4-1}
\usepackage[T1]{fontenc}
\usepackage[latin9]{inputenc}
\usepackage{geometry}
\usepackage{color}
\usepackage{babel}
\usepackage{amsmath}
\usepackage{amssymb}
\usepackage{graphicx}
\usepackage[unicode=true,pdfusetitle,
 bookmarks=true,bookmarksnumbered=false,bookmarksopen=false,
 breaklinks=false,pdfborder={0 0 0},pdfborderstyle={},backref=false,colorlinks=true]
 {hyperref}
\hypersetup{
 linkcolor=blue, urlcolor=blue, citecolor=blue}

\makeatletter
\usepackage{braket}
\usepackage{dsfont}
\usepackage{bm}
\date{}

\makeatother

\begin{document}
\title{Quantum thermodynamics aspects with a thermal reservoir based on $\mathcal{PT}$-symmetric
Hamiltonians}
\author{Jonas F. G. Santos}
\email{jonasfgs18@gmail.com}

\affiliation{Centro de Ciências Naturais e Humanas, Universidade Federal do ABC,
Avenida dos Estados 5001, 09210-580 Santo André, São Paulo, Brazil}
\author{Fabricio S. Luiz}
\email{fabriciosouzaluiz@gmail.com}

\affiliation{Faculdade de Ciências, Universidade Estadual Paulista, Rua Quirino
de Andrade 215,17033-360, Bauru, São Paulo, Brazil }
\begin{abstract}
We present results concerning aspects of quantum thermodynamics under
the background of non-Hermitian quantum mechanics for the dynamics
of a quantum harmonic oscillator. Since a better control over the
parameters in quantum thermodynamics processes is desired, we use
concepts from collisional model to introduce a simple prototype of
thermal reservoir based on $\mathcal{PT}$-symmetric Hamiltonians
and study its effects under the thermalization process of a single
harmonic oscillator prepared in a displaced thermal state. We verify
that controlling the $\mathcal{PT}$-symmetric features of the reservoir
allows to reverse the heat flow between system and reservoir, as well
as to preserve the coherence over a longer period of time and reduce
the entropy production. Furthermore, we considered a modified quantum
Otto cycle in which the standard hot thermal reservoir is replaced
by the thermal reservoir based on $\mathcal{PT}$-symmetric Hamiltonians.
By defining an effective temperature depending on the $\mathcal{PT}$-symmetric
parameter, it is possible to interchange the quantum Otto cycle configuration
from engine to refrigerator by varying the $\mathcal{PT}$-symmetric
parameter. Our results indicate that $\mathcal{PT}$-symmetric effects
could be useful to achieve an improvement in quantum thermodynamics
protocols such as coherence protection and entropy production reduction.
\end{abstract}
\maketitle

\section{Introduction}

The ability to control and manipulate quantum information is the paramount
key to develop the new generations of devices based on quantum properties.
Any quantum system is under the action of its surround which induces
the well-known decoherence phenomena on the system \cite{BreuerBook2002,Schlosshauer2007}.
Given that the coherence, for a specific basis, can be employed as
a resource for many protocols of quantum information and quantum communication
\cite{Winter2016,Dana2017}, the protection of coherence along a quantum
dynamics is an important task in modern devices, such as quantum computing
\cite{Marshall2019} and quantum cryptography \cite{Weekbrook2014}.
Furthermore, in many situations, as in quantum thermodynamics, the
system is allowed to change heat with a thermal environment. For example,
models of quantum heat engine and refrigerators employ two or more
thermal reservoirs to absorb or release heat from/on the system functioning
as working substance or refrigerant \cite{Chiara2020,Camati2019,Abah2020}.
Thus, the performance as well as the functioning of quantum thermal
machines depend crucially on how efficient is the quantum control
involved in thermodynamic protocols.

Given a general protocol in quantum information or thermodynamics,
one of the most important tasks is how to control the dynamics in
order to optimize the results. For unitary processes (where entropy
is kept unchanged), it is assumed a Hamiltonian representing an external
agent that drives the system from an initial to a final state. This
is particularly important in fluctuation theorems such as the Jarzynski
\cite{Jarzynski1997} and Crooks \cite{Crooks1999} relations. On
the other hand, in dissipative systems (where entropy changes), the
control may be implemented through some parameter of the thermal reservoir.
The entropy production during a thermalization, for instance, depends
fundamentally on the quantum aspects of the initial state, given that
some recent studies have shown that the entropy production is larger
for states with initial coherence \cite{Jader2019,Francica2019}.

In the last years, a new type of Hamiltonian control has arisen. Traditionally,
the standard quantum mechanics assumes as a bona-fide operator all
with the Hermiticity property, $O=O^{\dagger}$. This directly implies
a set of real eigenvalues, with a complete set of eigenstates that
are essential for defining a viable candidate for representing a physical
observable in quantum theory\cite{Ballantine} . However, since the
paper of Bender and Boettcher \cite{Bender1998}, it has been shown
that the set of Hamiltonians fulfilling the conditions of invariance
by spatial reflection (parity $\mathcal{P}$) and time reversal $\mathcal{T}$
also has real eigenvalues and thus it can represent physical systems.
This new condition to guarantee real spectra is known as $\mathcal{PT}$-symmetry,
and it gave rise to the non-Hermitian quantum mechanics \cite{Bender2015},
once $\mathcal{PT}$-symmetric Hamiltonians are not Hermitian in general.
Indeed, the range of interest in effects by considering systems described
by $\mathcal{PT}$-symmetric Hamiltonians is vast and includes quantum
optics and photonics \cite{R=0000A8uter2010,Regensburger2002}, time-dependent
Hamiltonians \cite{Zhang2019,Gong2013,Ponte2019}, applications to
non-commutative geometries \cite{Giri2009,Santos2018,Dey2012}, as
well as to fluctuation relations \cite{Deffner2015,Zeng2017,Gardas2016}.
Since the conditions for an operator to be a viable candidate to represent
a physical observable are that eigenvalues are real and that eingestates
are complete, the condition of orthogonality for a non-Hermitian $\mathcal{PT}$-symmetric
Hamiltonian $\mathcal{H}$ can be relaxed and substituted by a biorthogonality
\cite{Brody2014}, and the conection with the biorthogonal system
with the ortogonal is made by a nontrivial metric operator \cite{Mostafazadeh2004,Fring2016}.
Thus depending on the choice of the parameters associated to the metric
the effects in employing the Hermitian Hamiltonian to some protocol
can be considerably different \cite{Duarte2018,Dey2019}. For example,
Ref. \cite{Dey2019} has shown that the decoherence dynamics can be
modified changing the parameters of the metric. Another interesting
set of results concerns the study of quantum thermodynamic with non-Hermitian
Hamiltonian background. The pioneer paper in this direction is Ref.
\cite{Gardas2016}, in which the authors showed that the Jarzynski
equality and the Carnot bound, both expressing the second law of thermodynamics,
hold for non-Hermitian Hamiltonians.

In this work, we investigate some quantum thermodynamics aspects in
the context of non-Hermitian quantum mechanics. By introducing a particular
$\mathcal{PT}$- symmetric Hamiltonian and a given metric, we obtain
its Hermitian counterpart, from which we construct associated thermal
states. Using concepts of the collisional model theory \cite{Chiara2018,Ciccarello2017,Grimmer2016,Strasberg2017},
we write a Lindblad master equation where the system of interest is
a single harmonic oscillator and the thermal reservoir contains $\mathcal{PT}$-symmetric
features in the form of the parameter of the metric. We then consider
a thermalization process in which the initial state of the harmonic
oscillator is prepared in a displaced thermal state, such that it
is has an amount of coherence in the energy basis. We verify that
depending on the $\mathcal{PT}$-symmetric features of the thermal
reservoir, the heat flow between system and reservoir can be reversed,
the coherence can be preserved along the dynamics, as well as the
entropy production is considerably reduced. As a second part of our
work, we consider a modified quantum Otto cycle where the standard
hot thermal reservoir is replaced by a $\mathcal{PT}$-symmetric thermal
reservoir prototype. By writing explicit expressions for the net work
and for the heat exchanged with the cold and hot thermal reservoirs,
we verify that depending on the $\mathcal{PT}$-symmetric features,
the configuration of the Otto cycle can change from engine to refrigerator.

This work is organized as follows. In section \ref{sec:Review-on--symmetric}
we review some concepts on $\mathcal{PT}$-symmetry and non-Hermitian
quantum mechanics. Section \ref{sec:-symmetric-reservoir-via} is
dedicated to introduce the notion of $\mathcal{PT}$-symmetric thermal
reservoir via collisional model concepts as well as to derive the
thermal state with $\mathcal{PT}$-symmetric features. The thermalization
process of a system composed of a single harmonic oscillator prepared
in a displaced thermal state is analyzed in section \ref{sec:Reversing-the-heat},
where the heat, coherence and entropy production are quantified in
terms of $\mathcal{PT}$-symmetric feature of the thermal reservoir.
In section \ref{sec:-symmetric-reservoir-in} we study a modified
version of the quantum Otto cycle in which the hot thermal reservoir
is described by a $\mathcal{PT}$-symmetric thermal reservoir. Finally,
our conclusion and final remarks are drawn in section \ref{sec:Conclusion}.

\section{Review on $\mathcal{PT}$-symmetric Hamiltonian\label{sec:Review-on--symmetric}}

The standard quantum mechanics assumes that in order to a given operator
to be a viable candidate to represent a physical observable, it must
have real spectra and a set of complete eigenstates. Hermitian operators
fulfill these two conditions, in order that many textbooks assume
that any observable $O$ has to be Hermitian, $O=O^{\dagger}$\cite{Griffiths}.
However, Hermitian operators are not the only operator class that
has real spectra and a complete set of eigenvectors. As shown by Bender
and Boettcher \cite{Bender1998}, operators that are invariant under
parity $\mathcal{P}$ and time reversal $\mathcal{T}$ simultaneously
also have real spectra and a complete set of eigenvectors, being a
viable candidate to represent a physical observable. For example,
for a general $N$-dimensional Hamiltonian $\mathcal{H}\left(q_{\ell},p_{\ell}\right)$
with $\ell=1,...,N$, the reality of the spectrum is guaranteed by
the unbroken $\mathcal{P}\mathcal{T}$-symmetry \cite{Bender2002},
that can be translated into these two relations: $\left[\mathcal{H}\left(q_{\ell},p_{\ell}\right),\mathcal{P}\mathcal{T}\right]=0,$
and $\mathcal{P}\mathcal{T}\vert\Psi(t)\rangle=\vert\Psi(t)\rangle,$where
$\vert\Psi(t)\rangle$ is the eigenstate of $\mathcal{H}\left(q_{\ell},p_{\ell}\right)$.
These conditions mean that the Hamiltonian $\mathcal{H}\left(q_{\ell},p_{\ell}\right)$
must be invariant under the set of transformations \cite{Fring2007}

\begin{align}
\mathcal{P}\mathcal{T}q_{\ell}(\mathcal{P}\mathcal{T})^{-1} & \rightarrow-q_{\ell},\nonumber \\
\mathcal{P}\mathcal{T}p_{\ell}(\mathcal{P}\mathcal{T})^{-1} & \rightarrow p_{\ell},\nonumber \\
\mathcal{P}\mathcal{T}i(\mathcal{P}\mathcal{T})^{-1} & \rightarrow-i.\label{condition}
\end{align}

In the case of a non-Hermitian Hamiltonian $\mathcal{H}\left(q_{\ell},p_{\ell}\right)$
fulfilling the condition in Eq. (\ref{condition}), it is known as
$\mathcal{PT}$-symmetric Hamiltonian and the relation $\mathcal{H}\left(q_{\ell},p_{\ell}\right)=\mathcal{H}^{\mathcal{PT}}\left(q_{\ell},p_{\ell}\right)$
is achieved. The connection between the standard quantum mechanics
and the non-Hermitian quantum mechanics is performed by employing
the similarity transformation 

\begin{equation}
H\left(q_{\ell},p_{\ell}\right)=\eta\mathcal{H}\left(q_{\ell},p_{\ell}\right)\eta^{-1},\label{eq:relation1}
\end{equation}
 in which $\eta=\eta\left(q_{\ell},p_{\ell}\right)$ is the Dyson
map such that $\eta\eta^{-1}=\mathbb{I}$, with $\mathbb{I}$ the
identity operator \cite{Mostafazadeh2004,Fring2016,Luiz2020}. Furthermore,
using Eq. (\ref{eq:relation1}) and the Hermiticity relation of the
Hamiltonian $H\left(q_{\ell},p_{\ell}\right)$, it is easy to show
that the non-Hermitian Hamiltonian, $\mathcal{H}\left(q_{\ell},p_{\ell}\right)$,
satisfies the well-known quasi-Hermiticity relation $\Theta\mathcal{H}\left(q_{\ell},p_{\ell}\right)=\mathcal{H}^{\dagger}\left(q_{\ell},p_{\ell}\right)\Theta$,
where $\Theta=\eta^{\dagger}\eta$ is the metric operator to ensure
probability conservation \cite{Fring2016,Luiz2020}. Like the Hamiltonian,
non-Hermitian observables $\mathcal{O}$ are connected to Hermitian
observables through the similarity transformation, $O=\eta\mathcal{O}\eta^{-1}$.
This implies that the expected value for these observables is the
same, $\langle\phi(t)\vert O\vert\phi(t)\rangle=\langle\Psi(t)\vert\mathcal{O}\vert\Psi(t)\rangle$,
where $\vert\phi(t)\rangle=\eta^{-1}\vert\Psi(t)\rangle$\cite{Luiz2020}.
This allows us to obtain observables, choosing a similarity transformation
(or Dyson map operator) transforming our non-Hermitian Hamiltonian
into its Hermitian isospectral partner and working in the standard
quantum mechanics with this partner.

\section{Modeling a thermal reservoir through $\mathcal{PT}$- symmetric Hamiltonians\label{sec:-symmetric-reservoir-via}}

\begin{figure}
\includegraphics[scale=0.8]{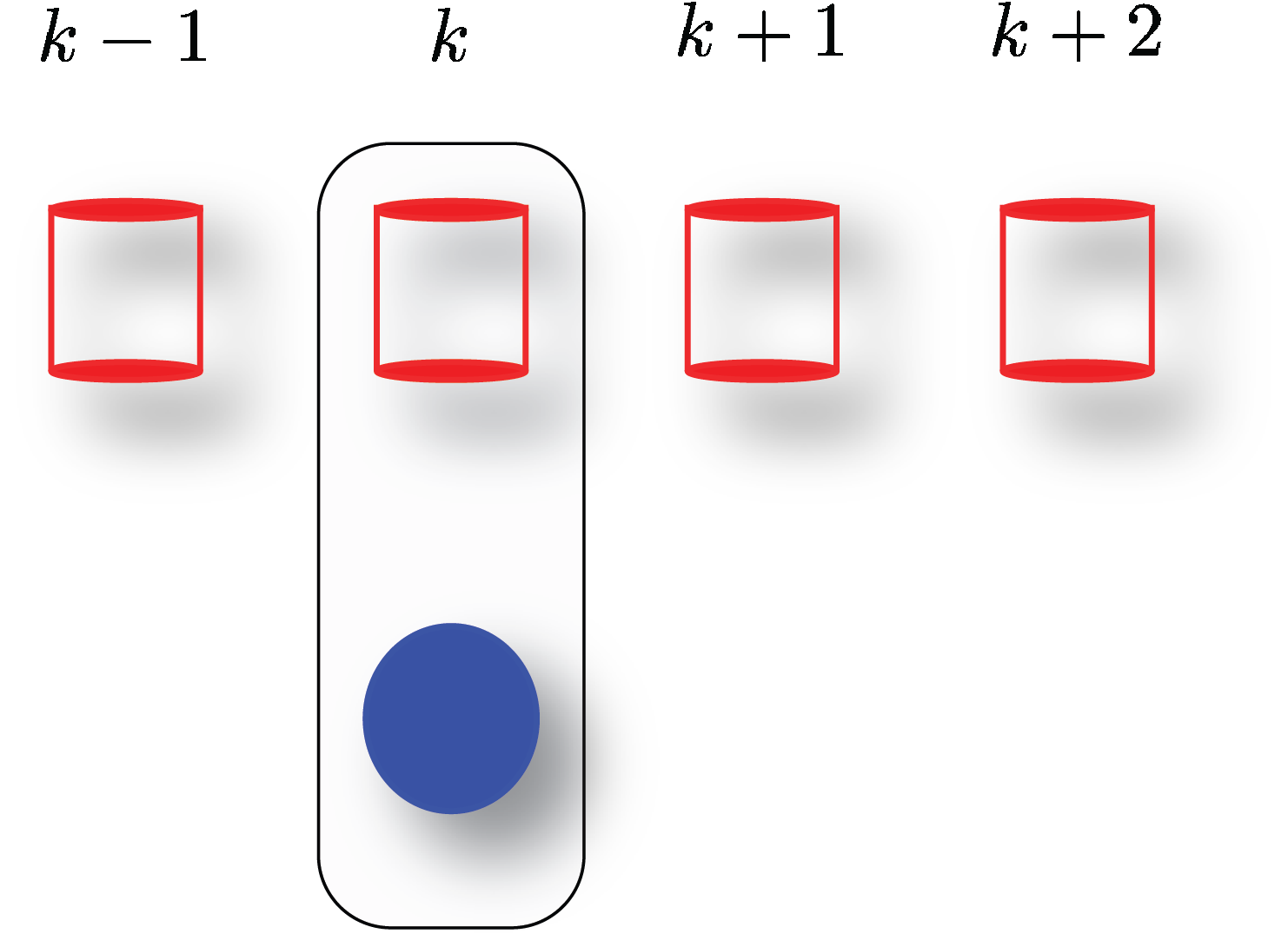}\caption{Illustration of the collisional model. The system (blue circle) interacts
with a particle of the thermal bath, which is composed of a large
number of non-interacting identical particles, all prepared in the
same thermal state. Once the system interacts with the particle $k$,
a new particle, $k+1$, is put to interact with the system.}

\label{Collisional model}
\end{figure}

In this section we use collisional model techniques to obtain the
master equation in which the thermal reservoir carriers $\mathcal{PT}$-symmetric
features. In order to review shortly the collisional model concepts,
one assumes a system $S$ in contact with a bath. In turn, the bath
is assumed to be a sufficient large collection of auxiliary systems
(ancilla), each of them prepared in the same thermal state. Besides,
the ancillas are considered to be non-interacting, a fact that is
associated to a Markovian dynamics of the system $S$ \cite{Ciccarello2017}.
Writing the initial state of the system and each ancilla as $\rho_{0}$
and $\rho^{\text{ancilla}}=\zeta^{\text{th}}\left(\bar{n}\right)$,
respectively, where $\bar{n}=\text{Tr\ensuremath{\left[a^{\dagger}a\rho^{\text{ancilla}}\right]}}$
is the average number of photons in a mode, with $a\left(a^{\dagger}\right)$
the annihilation (creation) operator, the joint state of the system
plus bath is given by $\sigma=\rho_{0}\otimes\left(\otimes_{j=1}^{N}\zeta_{j}^{\text{th}}\left(\bar{n}\right)\right)$.
In the collisional model, the system $S$ is assumed to interact with
just one ancilla at a time during a time interval $\delta t$. After
the collision with the ancilla $\zeta_{j}^{\text{th}}\left(\bar{n}\right)$,
it is discarded and a subsequent new ancilla $\zeta_{j+1}^{\text{th}}\left(\bar{n}\right)$
is brought to interact with the system. This process is repeated many
times. An illustration of the collisional model is depicted in Fig.
\ref{Collisional model}. Following Refs. \cite{Chiara2018,Ciccarello2017},
by assuming that the interaction time between the system and each
ancilla is sufficiently short, $\delta t\rightarrow0$, the collisional
model provides a Lindblad master equation \cite{Chiara2018}, which
can be given by

\begin{equation}
\frac{d\rho}{dt}=-i\left[H_{S},\rho\right]+\gamma\left(N+1\right)\mathcal{D}\left[a\right]\rho+\gamma N\mathcal{D}\left[a^{\dagger}\right]\rho,\label{ME01}
\end{equation}
with $H_{S}=\hbar\omega_{S}\left(a^{\dagger}a+1/2\right)$ the Hamiltonian
of the system, $\mathcal{D}[\mathcal{O}]=\mathcal{O}\rho\mathcal{O}-\frac{1}{2}\left(\rho\mathcal{O}^{\dagger}\mathcal{O}+\mathcal{O}^{\dagger}\mathcal{O}\rho\right)$
are the Lindblad operators, $N=\left(e^{\beta\hbar\omega}-1\right)^{-1}$
is the average number of photons associated to the bath, $\beta=1/k_{B}T$
is the inverse temperature, and $\gamma$ is the decay rate. For $N=0$,
Eq. (\ref{ME01}) represents pure loss.

As we will note throughout the work, all the states considered are
Gaussian \cite{Wang2007,Weedbrook2012,Adesso2014,SerafiniBook}. In
this situation, a suitable method to treat the thermalization dynamics
is employing the covariance matrix, defined as $\sigma_{ij}=\langle R_{i}R_{j}+R_{j}R_{i}\rangle_{\rho}-2\langle R_{i}\rangle_{\rho}\langle R_{j}\rangle_{\rho}$,
with the vector of operators $\vec{R}=\left(q,p\right)$, with $q\left(p\right)$
the position (momentum) operator. For a general Gaussian state $\rho=\rho\left(\vec{d},\sigma\right)$
with first moments $\vec{d}=\left(\langle q\rangle_{\rho},\langle p\rangle_{\rho}\right)$
and covariance matrix $\sigma$, the thermalization dynamics with
a Markovian reservoir is obtained from the two uncoupled differential
equations \cite{SerafiniBook},

\begin{eqnarray*}
\dot{\sigma} & = & \gamma\sigma+\gamma(2\bar{m}+1)\mathbb{I}_{2\times2},\\
\dot{\vec{d}} & = & -(\gamma/2)\vec{d},
\end{eqnarray*}
 with the following solutions

\begin{align}
\vec{d}\left(t\right) & =\vec{d}\left(0\right)e^{-\gamma t/2}\\
\sigma(t) & =\sigma\left(0\right)e^{-\gamma t}+\left(1+e^{-\gamma t}\right)\sigma_{\text{\text{aspt}}},
\end{align}
where $\sigma_{\text{\text{aspt}}}$ is the asymptotic covariance
matrix reached when $t\rightarrow\infty$ (complete thermalization).
This formalism also provides an expression to obtain the internal
energy of the system using the covariance matrix \cite{Santos2020},
given by $\mathcal{U}_{t}=\text{\ensuremath{\hbar\omega}Tr}\left[\sigma(t)\right]/4$.

We now pass to consider a $\mathcal{PT}$-symmetric Hamiltonian following
the concepts presented in Sec. \ref{sec:Review-on--symmetric} to
construct thermal states for the ancillas of the bath. Let us assume
the following $\mathcal{PT}$-symmetric Hamiltonian given by \cite{Dey2019}

\begin{equation}
\mathcal{H}^{\mathcal{PT}}=\frac{p^{2}}{2m}+\frac{1}{2}m\omega^{2}q^{2}+2i\omega\epsilon pq,\label{PT01}
\end{equation}
which has a real spectra. A similar structure for the $\mathcal{PT}$-symmetric
Hamiltonian was employed in Ref. \cite{Dey2019} to claim that $\mathcal{PT}$-symmetry
can be useful to control the decoherence of a two-level system. For
simplicity we assume the following expression for the Dyson operator,
$\eta=e^{\frac{\epsilon}{m\omega\hbar}p^{2}}$. As we will see, this
metric is enough to elucidate all relevant results in the present
work. In performing the Hermitization procedure as in Sec. \ref{sec:Review-on--symmetric}
we obtain the Hermitian Hamiltonian

\begin{align}
H & =\eta\mathcal{H}^{\mathcal{PT}}\eta^{-1}\nonumber \\
 & =\frac{\mu^{2}}{2m}p^{2}+\frac{1}{2}m\omega^{2}q^{2}+\hbar\omega\epsilon,\label{Hermitian01}
\end{align}
with $\mu^{2}=\left(1+4\epsilon^{2}\right)$. Hamiltonian in Eq. (\ref{Hermitian01})
is clearly that of a one-mode harmonic oscillator with a shift of
energy. Thus, it is possible to construct a thermal state which will
be the ancillary states of the bath. A thermal state can be written
in general as $\zeta^{\text{th}}=e^{-\beta H}/\mathcal{Z}$, where
$\mathcal{Z}=\sum_{n}e^{-\beta H}$ is the partition function. By
using Eq. (\ref{Hermitian01}), the associated thermal state reads

\begin{equation}
\zeta^{\text{th}}=\frac{e^{-\beta\hbar\omega\mu\left(n+1/2\right)}}{\sum_{n}e^{-\beta\hbar\omega\mu\left(n+1/2\right)}}.\label{Thermal01}
\end{equation}

Writing the thermal state in Eq. (\ref{Thermal01}) in the Fock basis
one has

\begin{equation}
\zeta^{\text{th}}\left(\bar{m}\right)=\sum_{n}\frac{\bar{m}^{n}}{\left(\bar{m}+1\right)^{n+1}}|n\rangle\langle n|,\label{thermal02}
\end{equation}
with $\bar{m}=\left(e^{\beta\hbar\omega\mu}-1\right)^{-1}$. Throughout
the rest of the work we assume $N=\bar{m}$, meaning that the thermal
states of the ancilla in the bath are given by Eq. (\ref{thermal02}).
This implies that the master equation in Eq. (\ref{ME01}) shall also
includes the effects of the $\mathcal{PT}$-symmetric Hamiltonian
which physically is included by the factor $\mu$. In order to elucidate
the physical meaning of $\mathcal{PT}$-symmetric features in the
thermal reservoir prototype, we observe from the average number of
photons $\bar{m}$ that it is possible to define an effective temperature
as $\beta^{\text{eff}}=\left(1+4\epsilon^{2}\right)\beta$, where
obviously for $\epsilon=0$ we have $\beta^{\text{eff}}=\beta$. In
order to be clear, in the following we use the expression ``$\mathcal{PT}$-symmetric
thermal reservoir'' as being that modeled with ancillas prepared
in thermal states using $\mathcal{PT}$-symmetric Hamiltonians.

\section{Reversing the heat flow and protecting coherence using a $\mathcal{PT}$-Symmetric
thermal reservoir\label{sec:Reversing-the-heat}}

In order to investigate how a thermal reservoir carrying $\mathcal{PT}$-symmetric
features could influence the thermalization dynamics of a single-mode
Gaussian state, we consider an example of initial state with coherence
in the Fock basis. The motivation for this is that we are able to
study not only the heat exchanged between system and reservoir but
also the coherence dynamics introduced by the $\mathcal{PT}$-symmetric
features of the bath.

The system under consideration is a single harmonic oscillator. The
most paradigmatic Gaussian state associated to the Hamiltonian of
the system is a thermal state $\zeta^{th}(\bar{n})$. To introduce
coherence on the state $\zeta^{th}(\bar{n})$, we use the mechanism
consisting in applying the displacement operator $D\left(\alpha\right)=e^{\alpha a^{\dagger}-\alpha^{\ast}a}$,
such that the state $\rho=D\left(\alpha\right)\zeta^{th}(\bar{n})D\left(\alpha\right)^{\dagger}$
possesses an amount of coherence.

To quantify the coherence of a Gaussian state, we use an entropic
quantifier based on the relative entropy $S\left(\sigma_{1}||\sigma_{2}\right)=\sigma_{1}\text{ln\ensuremath{\left[\sigma_{1}\right]}-\ensuremath{\sigma_{1}\text{ln\ensuremath{\left[\sigma_{2}\right]}}}}$\cite{Baumgratz2014}.
In Ref. \cite{Xu2016} the same measure is particularized for Gaussian
states, such that one way of quantifying coherence of a general $N$-mode
Gaussian state $\rho=\rho(\vec{d},\sigma)$ is given by

\begin{equation}
C(\rho)=S(\zeta^{\text{th}})-S(\rho),\label{coherencemeasure}
\end{equation}
 where $S(\cdot)$ is the von Neumann entropy,

\begin{equation}
S(\rho)=-\sum_{j=1}^{N}\left[\frac{\nu_{j}-1}{2}\ln\left(\frac{\nu_{j}-1}{2}\right)-\frac{\nu_{j}+1}{2}\ln\left(\frac{\nu_{j}+1}{2}\right)\right],
\end{equation}
with $\left\{ \nu_{j}\right\} _{j=1}^{N}$ the symplectic eigenvalues
of $\sigma$, and $\zeta^{\text{th}}$ is a $N$-mode reference thermal
state with average number of photons $\left\{ \bar{k}_{j}\right\} _{j=1}^{N}$written
in terms of the first moments and the covariance matrix of $\rho$
and given by \cite{Xu2016}

\begin{equation}
\bar{k}_{j}=\frac{1}{4}\left[\sigma_{11}^{j}+\sigma_{22}^{j}+\left(d_{1}^{j}\right)^{2}+\left(d_{2}^{j}\right)^{2}-2\right].
\end{equation}

\begin{figure}
\includegraphics[scale=0.8]{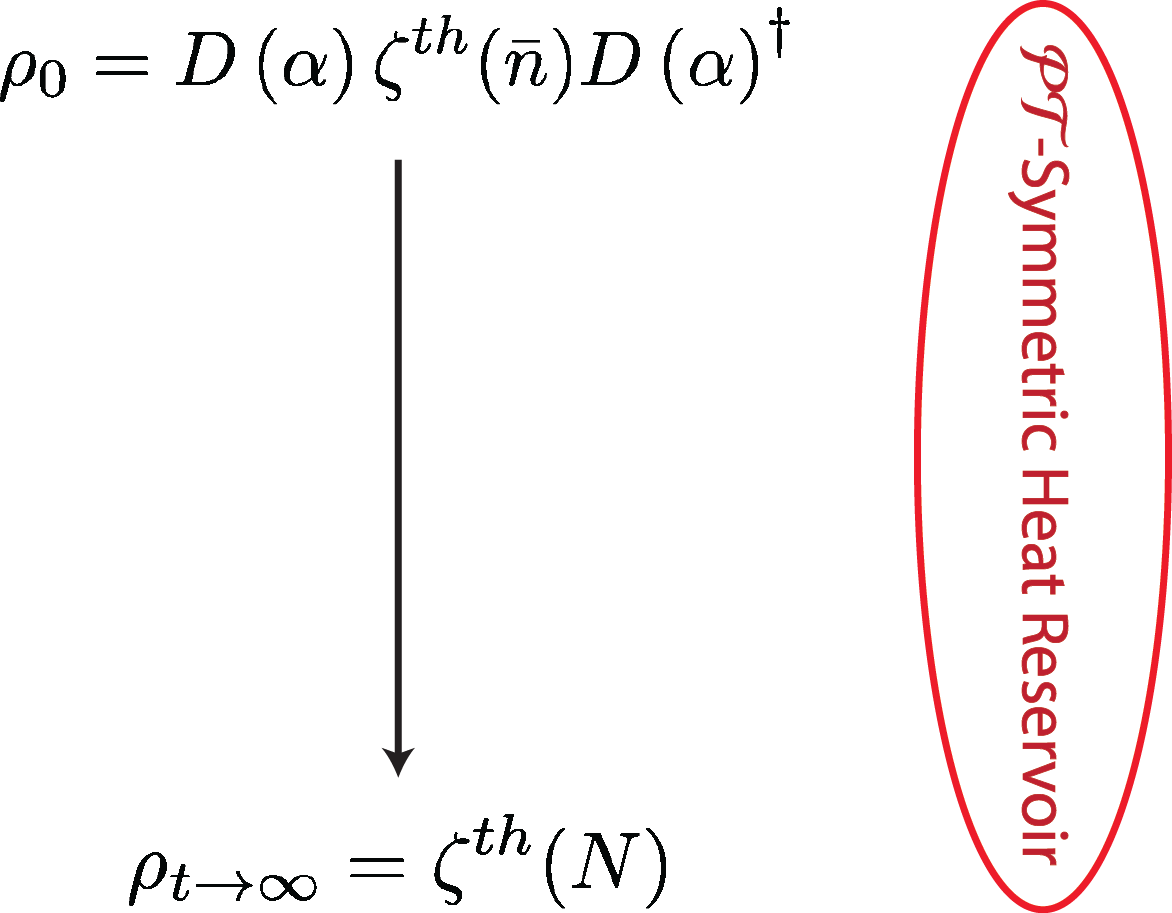}\caption{An illustration of the thermalization process of a single harmonic
oscillator prepared in a displaced thermal state with average number
of photons $\bar{n}$ interacting with the $\mathcal{PT}$-symmetric
thermal reservoir with effective temperature $\beta^{\text{eff}}=\left(1+4\tau^{2}\right)\beta$.}

\label{Thermalization}
\end{figure}

The coherence quantifier in Eq. (\ref{coherencemeasure}) satisfies
the following properties: $C(\rho)\geq0$, $C(\rho)=0$ if and only
if $\rho$ is a tensor product of thermal states, and $C(\rho)\geq C(\Phi_{IGC}\rho)$,
where $\Phi_{IGC}$ is an incoherent Gaussian channel \cite{Xu2016}.
An illustration of the thermalization protocol is depicted in Fig.\ref{Thermalization}.

The heat exchanged between system and thermal reservoir can be written
in terms of the covariance matrix as

\begin{align}
\langle Q\rangle & =\frac{\hbar\omega}{4}\left(\text{Tr}\left[\sigma\left(t\right)\right]-\text{Tr}\left[\sigma\left(0\right)\right]\right)\label{heat}\\
 & =\hbar\omega\left(N-\bar{n}\right)\left(1-e^{-\gamma t}\right),
\end{align}
with $N=\left(e^{\beta\hbar\omega\mu}-1\right)^{-1}$. Another important
quantity during a thermalization is the entropy production, which
can be written as

\begin{equation}
\langle\Sigma\rangle=-\beta\Delta\mathcal{U}_{\tau_{2},\tau_{1}}+\Delta S_{\tau_{2},\tau_{1}},
\end{equation}
where $\Delta\mathcal{U}_{\tau_{2},\tau_{1}}=\mathcal{U}_{\tau_{2}}-\mathcal{U}_{\tau_{1}}$
is the variation of internal energy associated to the thermalization
process and $\Delta S_{\tau_{2},\tau_{1}}=S(\rho_{\tau_{2}})-S(\rho_{\tau_{1}})$.

\begin{figure}
\includegraphics[scale=0.8]{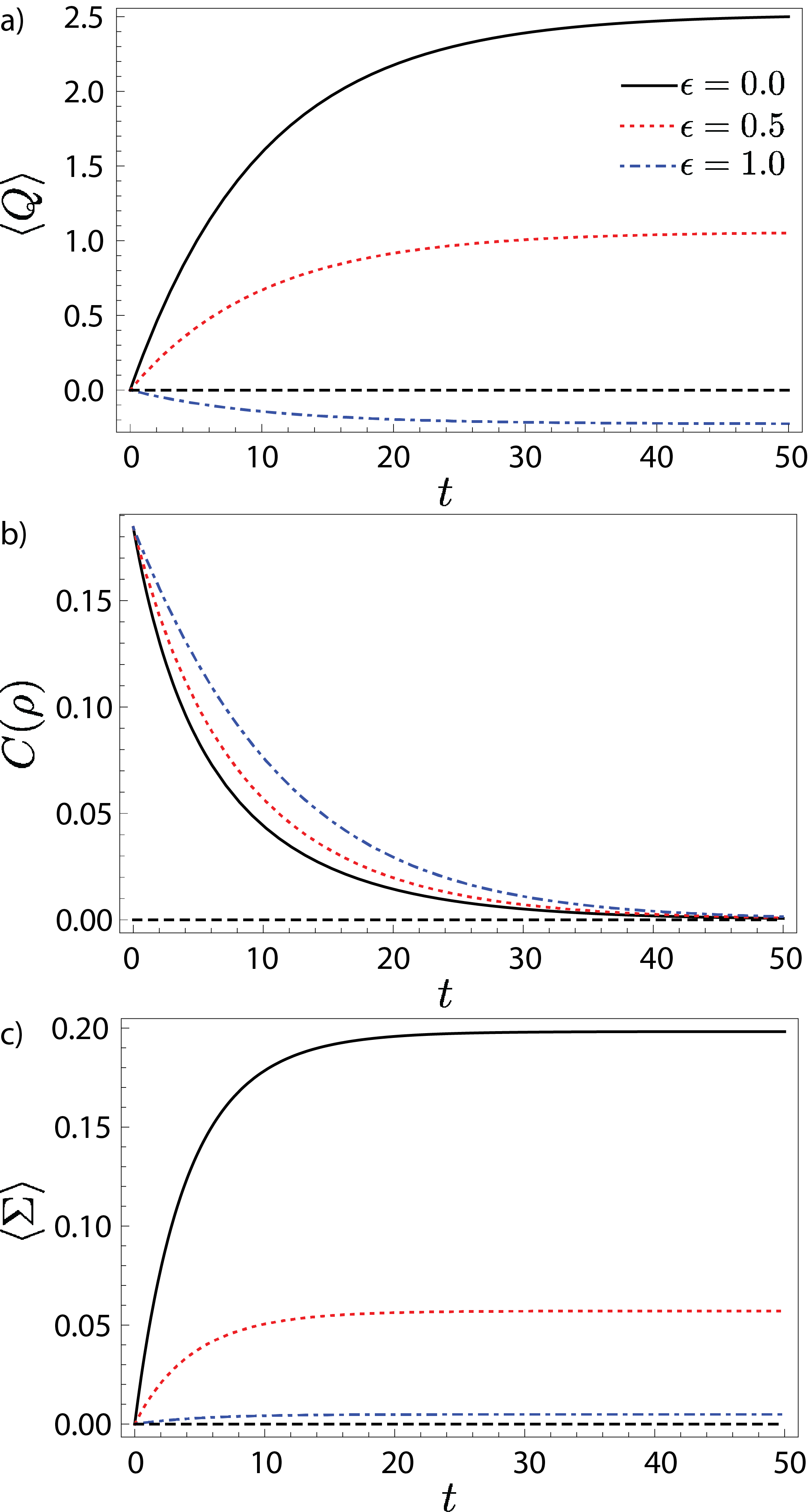}

\caption{Thermodynamic quantities as a function of the thermalization time
with the $\mathcal{PT}$-symmetric thermal reservoir. A single-mode
displaced thermal state with average number of photons $\bar{n}$
is prepared to interact with a thermal reservoir with average number
of photons $N$ and effective temperature $\beta^{\text{eff}}=\left(1+4\epsilon^{2}\right)\beta$.
a) Heat exchanged between system and thermal reservoir, b) coherence
dynamics during the thermalization, and c) entropy production during
the thermalization. We consider three cases: $\epsilon=0$ (solid
black lines), $\epsilon=0.5$ (dotted red lines), and $\epsilon=1.0$
(dashed dotted blue lines). We consider a set of parameters for the
thermal bath such that $\beta\hbar\omega=0.2$, with a decay rate
$\gamma=0.1$. The action of the displacement operator is such that
the initial state is located $\left(q_{0},p_{0}\right)=\left(1,1\right)$
on the phase-space.}

\label{heat_coherence_EP_A}
\end{figure}

Figure \ref{heat_coherence_EP_A} presents the heat exchanged, coherence,
and the entropy production as a function of the thermalization time
$t$ with the $\mathcal{PT}$-symmetric thermal reservoir. The initial
state is assumed to be a displaced thermal state $\rho=D\left(\alpha\right)\zeta^{th}(\bar{n})D\left(\alpha\right)^{\dagger}$,
with an average number of photons $\bar{n}=2$, and $\alpha$ such
that the initial position of the state on the phase-space is $\left(q_{0},p_{0}\right)=\left(1,1\right)$.
For the thermal reservoir, it was considered an effective temperature
$\beta^{\text{eff}}=\left(1+4\epsilon^{2}\right)\beta$, with $\beta\hbar\omega=0.2$.
To see how $\mathcal{PT}$-symmetric features of the thermal bath
influence the thermalization dynamics we have set the value of $\epsilon$
to be $0$ (solid black lines), $0.5$ (dotted red lines), and $1.0$
(dashed dotted blue lines). The same results would be verified in
the case of a squeezed thermal state as initial one-mode. In considering
these two class of initial states, we are encompassing all the mechanics
to generate coherence in a single-mode Gaussian state.

For $\epsilon=0$ we recover the standard thermal bath. Figure \ref{heat_coherence_EP_A}-a)
shows the heat exchanged between system and thermal bath as a function
of the thermalization time. It can be observed that, depending on
the value of the $\mathcal{PT}$-symmetric parameter $\tau$, the
heat flow can be reversed, passing to flow from the reservoir to the
system. This is physically explained by looking at the effective temperature
$\beta^{\text{eff}}=\left(1+4\epsilon^{2}\right)\beta$ which changes
for different values of $\epsilon$. In Fig. \ref{heat_coherence_EP_A}-b)
we present the coherence as a function of the thermalization time.
The results shows that increasing the value of $\epsilon$, i.e.,
becoming the $\mathcal{PT}$-symmetric feature more intensive, it
is possible to effectively protect the coherence of the initial state
from the decoherence effects due the thermalization. Finally, Figure
\ref{heat_coherence_EP_A}-c) depicts the entropy production as a
function of the thermalization time, where it can be noted that the
entropy production is reduced when $\epsilon$ is increased.

Some of the results presented in Fig. \ref{heat_coherence_EP_A} have
been discussed in previous papers. In Ref. \cite{Dey2019} the authors
showed that the decoherence dynamics is modified if the environment
is assumed to carry non-Hermitian signatures. Furthermore, Ref. \cite{Duarte2018}
exploits a $\mathcal{PT}$-symmetric interaction Hamiltonian in the
linear response theory scenario and argues that depending on the $\mathcal{PT}$-symmetric
properties of the interaction, there could be have a change in the
heat flow from the system to the reservoir as well as a coherence
protection of the initial state. Our results indicate that the same
effects could be obtained by considering a simple model of $\mathcal{PT}$-symmetric
thermal reservoir. Besides, the introduction of an effective temperature
shows physically how $\mathcal{PT}$-symmetry affects the dynamics
of the system.

\section{$\mathcal{PT}$-symmetric thermal reservoir in a quantum Otto cycle
\label{sec:-symmetric-reservoir-in}}

\begin{figure}
\includegraphics[scale=0.7]{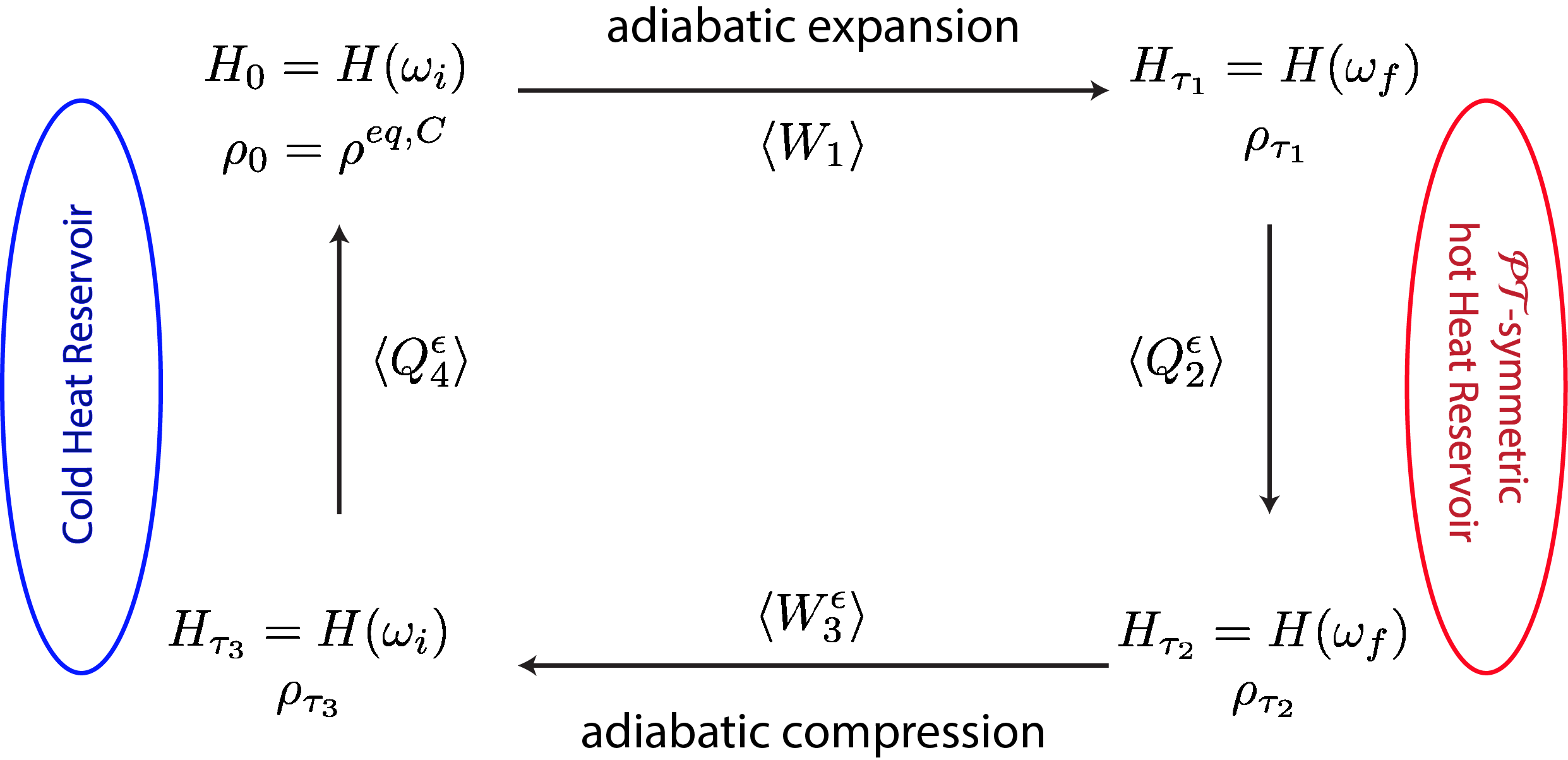}

\caption{Modified quantum Otto cycle. The working substance is a single harmonic
oscillator, with the standard hot thermal reservoir replaced by the
$\mathcal{PT}$-symmetric thermal reservoir prototype. The unitary
strokes are performed quasistatically whereas the two thermalization
strokes are assumed to be complete. These aspects do not affects our
results and they are valid for a finite-time regime. Depending on
the $\mathcal{PT}$-symmetric features of the hot thermal reservoir
the quantum Otto cycle can works as a engine or as a refrigerator.}

\label{cycle}
\end{figure}

Motivated by the results in the last section, in particular the reversing
of the heat flow between system and thermal reservoir, here we employ
the model of $\mathcal{PT}$-symmetric thermal reservoir in a modified
quantum Otto cycle to observe its effects. We assume the simpler form
for the quantum Otto cycle, i.e., the unitary strokes are performed
quasistatically and the thermalization with the hot and cold reservoirs
are complete. These restrictions are not relevant in our discussion,
once the $\mathcal{PT}$-symmetric parameter $\epsilon$ introduces
just an effective temperature $\beta^{\text{eff}}$ on the thermal
reservoir. Thus, taking in account a finite-time quantum Otto cycle
will not bring any advantage in our discussion. The illustration of
the cycle is shown in Fig. \ref{cycle}. The hot thermal reservoir,
with inverse temperature $\beta_{\text{hot}}$, is assumed to have
$\mathcal{PT}$-symmetric features in an exact way as in the Sec.
\ref{sec:Reversing-the-heat}, whereas the cold thermal reservoir
is assumed to be purely thermal, i.e., it is a standard thermal bath
with inverse temperature $\beta_{\text{cold}}>\beta_{\text{hot}}$.
The quantum Otto cycle is fueled by a harmonic oscillator, prepared
initially in thermal equilibrium with the cold thermal reservoir with
state denoted by $\rho_{0}=\zeta_{\text{cold}}^{\text{eq}}$. The
quantum Otto cycle is described as follows.

First stroke. The working substance is submitted to a unitary process
performed quasistatically such that the final state is given $\rho_{\tau_{1}}=\zeta_{\text{cold}}^{\text{eq}}$
and the frequency of the Hamiltonian is changed from $\omega_{i}$
to $\omega_{f}$. The work performed in this stage is given by $\langle W_{1}\rangle=\left(\hbar\omega_{f}\text{Tr\ensuremath{\left[\sigma_{\tau_{1}}\right]}}-\hbar\omega_{i}\text{Tr\ensuremath{\left[\sigma_{0}\right]}}\right)/4$.

Second stroke. A complete thermalization between the working substance
and the hot thermal reservoir is implemented, such that the final
state is given by $\rho_{\tau_{2}}=\zeta_{\text{hot}}^{\text{eq,\ensuremath{\epsilon}}}$.
Note that the inclusion of the parameter $\epsilon$ in the thermal
state after the thermalization with the hot thermal reservoir means
that the effective temperature of $\rho_{\tau_{2}}$ depends on $\epsilon$.
The frequency of the Hamiltonian is kept unchanged during this process,
with heat exchanged given by $\langle Q_{2}^{\epsilon}\rangle=\hbar\omega_{f}\left(\text{Tr\ensuremath{\left[\sigma_{\tau_{2}}\right]}}-\text{Tr\ensuremath{\left[\sigma_{\tau_{1}}\right]}}\right)/4.$

Third stroke. The frequency of the Hamiltonian is quasistatically
changed back to $\omega_{i}$, such that the working substance state
at the final is $\rho_{\tau_{3}}=\zeta_{\text{hot}}^{\text{eq,\ensuremath{\epsilon}}}$.
The work performed in this stroke is $\langle W_{3}^{\epsilon}\rangle=\left(\hbar\omega_{i}\text{Tr\ensuremath{\left[\sigma_{\tau_{3}}\right]}}-\hbar\omega_{f}\text{Tr\ensuremath{\left[\sigma_{\tau_{2}}\right]}}\right)/4$.

Fourth stroke. Finally, the working substance completely thermalizes
with the cold reservoir, closing the cycle, such that $\rho_{\tau_{4}}=\zeta_{\text{cold}}^{\text{eq}}$.
Again, the frequency is kept unchanged. The heat exchanged with the
cold thermal reservoir is $\langle Q_{4}^{\epsilon}\rangle=\hbar\omega_{i}\left(\text{Tr\ensuremath{\left[\sigma_{0}\right]}}-\text{Tr\ensuremath{\left[\sigma_{\tau_{3}}\right]}}\right)/4$.

In a quantum Otto cycle, depending on the thermodynamic quantities,
i.e., the net work $\langle W_{\text{net}}^{\epsilon}\rangle=\langle W_{1}\rangle+\langle W_{3}^{\epsilon}\rangle$,
as well as the two heat exchanged $\langle Q_{2}^{\epsilon}\rangle$
and $\langle Q_{4}^{\epsilon}\rangle$, it works as engine or refrigerator.
For the former, the conditions are $\langle W_{\text{net}}^{\epsilon}\rangle<0,$
$\langle Q_{2}^{\epsilon}\rangle>0$, and $\langle Q_{4}^{\epsilon}\rangle<0$,
whereas for the latter, $\langle W_{\text{net}}^{\epsilon}\rangle>0,$
$\langle Q_{2}^{\epsilon}\rangle<0$, and $\langle Q_{4}^{\epsilon}\rangle>0$.
The thermodynamic quantities can be written explicitly as

\begin{align*}
\langle Q_{2}^{\epsilon}\rangle & =\frac{\hbar\omega_{f}}{2}\left\{ \coth\left[\frac{\hbar\omega_{f}\beta_{\text{hot}}^{\text{eff}}}{2}\right]-\coth\left[\frac{\hbar\omega_{i}\beta_{\text{cold}}}{2}\right]\right\} ,\\
\langle Q_{4}^{\epsilon}\rangle & =\frac{\hbar\omega_{i}}{2}\left\{ \coth\left[\frac{\hbar\omega_{i}\beta_{\text{cold}}}{2}\right]-\coth\left[\frac{\hbar\omega_{f}\beta_{\text{hot}}^{\text{eff}}}{2}\right]\right\} ,\\
\langle W_{\text{net}}^{\epsilon}\rangle & =-\frac{\hbar\left(\omega_{f}-\omega_{i}\right)}{2}\left\{ \coth\left[\frac{\hbar\omega_{f}\beta_{\text{hot}}^{\text{eff}}}{2}\right]-\coth\left[\frac{\hbar\omega_{i}\beta_{\text{cold}}}{2}\right]\right\} ,
\end{align*}
where $\beta_{\text{hot}}^{\text{eff}}=\sqrt{1+\epsilon^{2}}\beta_{\text{hot}}$
and the conditions $\omega_{f}/\omega_{i}<\beta_{\text{cold}}/\beta_{\text{hot}}^{\text{eff}}$
and $\omega_{f}/\omega_{i}>\beta_{\text{cold}}/\beta_{\text{hot}}^{\text{eff}}$
have to be fulfilled for the engine and refrigerator configuration,
respectively.

\begin{figure}
\includegraphics[scale=0.9]{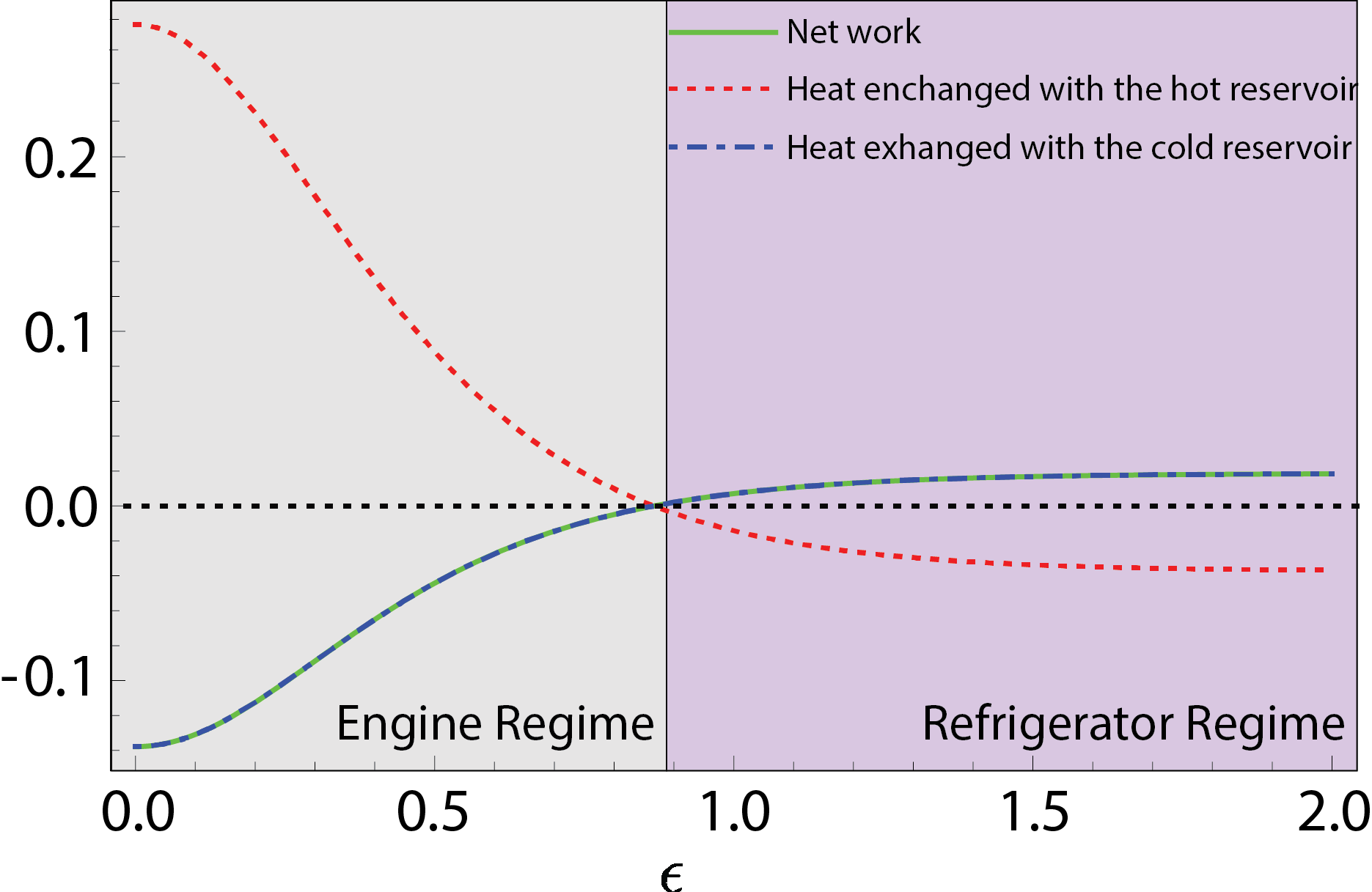}

\caption{Thermodynamic quantities for the modified quantum Otto cycle as a
function of the $\mathcal{PT}$-symmetric parameter $\epsilon$. Net
work (solid green line), heat exchanged with the hot reservoir (dotted
red line) and heat exchanged with the cold reservoir (dotted dashed
line). We set the parameters of the cycle to be $\omega_{f}=2\omega_{i}$
and $\beta_{\text{cold}}=4\beta_{\text{hot}}$.}

\label{ThermQuanti}
\end{figure}

For the quantum Otto cycle operating as an engine the efficiency is
given by $\eta=-\langle W_{\text{net}}^{\tau}\rangle/\langle Q_{2}^{\tau}\rangle=1-\omega_{i}/\omega_{f}$,
while operating as a refrigerator the coefficient of performance is
$\text{COP}=\langle Q_{4}^{\tau}\rangle/\langle W_{\text{net}}^{\tau}\rangle=\omega_{i}/\left(\omega_{f}-\omega_{i}\right)$.
Thus, the inclusion of $\mathcal{PT}$-symmetric features in the hot
thermal reservoir does not affect the performance of the quantum Otto
cycle. It must be stressed that these results would be obtained even
considering a finite-time regime and partial thermalizations \cite{Camati2019}.
However, from the Sec. \ref{sec:Reversing-the-heat}, depending on
the value of $\epsilon$ it is possible to change the direction of
the heat flow between the system and hot thermal reservoir. This fact
implies that we can move from the engine to refrigerator configurations
and vice-versa by controlling the $\mathcal{PT}$-symmetric parameter
$\epsilon$. Figure \ref{ThermQuanti} depicts exactly these results
by showing the net work, and the hot and cold heat exchanged as a
function of the parameter $\epsilon$. As it can be observed from
Fig. \ref{ThermQuanti}, for a critical value of $\epsilon$, the
quantum cycle configuration changes from engine to refrigerator. The
exact value of $\epsilon_{\text{c}}$ is given by

\begin{equation}
\epsilon_{\text{c}}=\frac{\sqrt{\left(\omega_{i}\beta_{\text{cold}}\right)^{2}-\left(\omega_{f}\beta_{\text{hot}}\right)^{2}}}{2\omega_{f}\beta_{\text{hot}}}.
\end{equation}

This result is the main concerning application of $\mathcal{PT}$-symmetric
effects in quantum Otto cycle and in principle could be tested in
quantum machines based on optical devices \cite{Passos2019}.

It could be enlightening to conclude this section by comparing our
model of thermal reservoir, which is performed by assuming $\mathcal{PT}$-
symmetric Hamiltonians for the ancillas, with some other models proposed
in the literature. As showed above, the inclusion of the $\mathcal{PT}$-
symmetric property and consequently the Dyson map allowed to add another
parameter in the thermal reservoir besides the temperature, resulting
in the behaviors as reversing of the heat flow and changing the regime
in a quantum Otto cycle from engine to refrigerator. Other proposals,
however, have also considered non-classical effects in thermal reservoir
models to obtain advantages. A well-known example is that of a squeezed
thermal reservoir \cite{Manzano2018}, employed to derive a generalized
Carnot bound \cite{Ro=0000DFnagel2014} as well as in quantum Otto
cycles \cite{Klaers2017,Assis2020,Manzano2016}. In this case, it
is also possible to write an effective temperature depending on the
squeezing parameter $r$, given by $T_{\text{eff}}=T_{0}\cosh\left(2r\right)$
\cite{Klaers2017}. A similar effect is obtained here, though with
a different mathematical expression. Another situation happens when
the ancillas of the thermal reservoir contain an amount of coherence.
In Ref. \cite{Scully2003} the authors showed that, despite that in
the high-temperature limit the quantum coherence becomes small enough
and the bath is essentially thermal, the phase associated to the coherence
may also be employed to define an effective temperature. Furthermore,
quantum coherence in a non-equilibrium reservoir is also employed
to convert heat in ordered energy in in the system \cite{Rodrigues2019}.
The main different aspect between the present work and the other proposals
is that the introduction of an effective temperature comes from a
mathematical property of non-Hermitian operators, i.e., the $\mathcal{PT}$-symmetry
of the Hamiltonian operator. 

\section{Conclusion\label{sec:Conclusion}}

The emergent interest in considering non-Hermitian Hamiltonians with
real spectra, i.e, fulfilling $\mathcal{PT}$-symmetric conditions,
in different branches of quantum physics have been evidenced by considerable
theoretical and experimental developments. The powerful of modeling
quantum systems with $\mathcal{PT}$-symmetric features could be useful
for future quantum devices. In this work, we have investigated some
quantum thermodynamics aspects in the scenario in which the thermal
reservoir is modeled such that it carries $\mathcal{PT}$-symmetric
effects. Employing concepts from the collisional model theory we write
a Lindblad master equation which governs the thermalization dynamics
of a single harmonic oscillator interacting with a $\mathcal{PT}$-symmetric
thermal reservoir. It was possible to write an effective temperature
for the thermal reservoir, $\beta_{\text{hot}}^{\text{eff}}=\sqrt{1+\epsilon^{2}}\beta_{\text{hot}}$
, which includes $\mathcal{PT}$-symmetric effects.

The first set of results concerns in studying the thermalization dynamics
of a single harmonic oscillator prepared in a displaced thermal state.
We verified that depending on the $\mathcal{PT}$-symmetric features
of the thermal reservoir, the heat flow can be reversed between system
and reservoir, the coherence is preserved over a longer period of
time, as well as the entropy production is considerably reduced. As
the second part of the work, we considered a modified quantum Otto
cycle, where the standard hot thermal reservoir was replaced by a
$\mathcal{PT}$-symmetric thermal reservoir. Although the performance
of the Otto cycle does not depend on $\mathcal{PT}$-symmetric features,
we showed that varying the $\mathcal{PT}$-symmetric parameter of
the hot thermal reservoir implies in changing the configuration of
the cycle from engine to refrigerator.

The results goes in the direction of recent contributions showing
the relevance in considering non-Hermitian Hamiltonians in quantum
protocols. We hope that this work can help to unveil the role played
by $\mathcal{PT}$-symmetric Hamiltonians in quantum thermodynamics.

\section*{Acknowledgments}

Jonas F. G. Santos acknowledges São Paulo Research Grant No. 2019/04184-5
and Federal University of ABC for support. Fabricio S. Luiz kindly
acknowledges National Council for Scientific and Technological Develompment
(CNPq) Research Grant No. 151435/2020-0 and São Paulo State University
for support.

\end{document}